\documentclass[prl,aps,twocolumn]{revtex4}

\usepackage[dvips]{graphicx}
\usepackage{epsfig}

\begin{document}

\title{Suppression of timing errors
    in short overdamped Josephson junctions}
\author{Andrey L. Pankratov$^1$ and Bernardo Spagnolo$^2$}
\email[]{alp@ipm.sci-nnov.ru}
 \email[]{spagnolo@unipa.it}
\homepage[]{http://gip.dft.unipa.it}
\affiliation{$^1$ Institute for
Physics of Microstructures of RAS, GSP-105, Nizhny Novgorod, 603950, Russia
\\$^2$ INFM and Dipartimento di Fisica e Tecnologie Relative, Group of Interdisciplinary Physics,\\
Universit$\grave{a}$ di Palermo, Viale delle Scienze, I-90128
Palermo, Italy}

\begin{abstract}
    The influence of fluctuations and periodical driving on temporal
    characteristics of short overdamped Josephson junction is analyzed.
    We obtain the standard deviation of the switching
time in the presence of a dichotomous driving force for arbitrary
noise intensity and in the frequency range of practical interest.
For sinusoidal driving the resonant activation effect has been
observed. The mean switching time and its standard deviation have
a minimum as a function of driving frequency. As a consequence the
optimization of the system for fast operation will simultaneously
lead to minimization of timing errors.
\end{abstract}

\maketitle


 The Rapid Single Flux Quantum (RSFQ) electronic devices
are promising candidates to built a petaflop computer due to high
operating frequencies of RSFQ elements close to 1 THz \cite{rsfq}.
Moreover they are of particular interest in solid-state quantum
information processing. They may be used both for realization of
qubits and for characterization of a macroscopic quantum behavior,
e.g., readout electronics for quantum computing \cite{MSS-RF}. The
processes going on in RSFQ devices are based on a reproduction of
quantum pulses due to spasmodic changing by $2\pi$ of the phase
difference of damped Josephson junctions (JJ). The major
restriction in the development of RSFQ logic circuits is given by
influence of fluctuations \cite{rsfq,lik-B,rl,s}. Moreover, the
operating temperatures of the high-\emph{$T_c$} superconductors
lead to higher noise levels by increasing the probability of
associated errors. Recently lots of investigations were performed
to study pulse jitter and timing errors in RSFQ circuits
\cite{rl,s}. The Timing errors is one of reasons
 limiting a 16-bit RSFQ microprocessor prototype
 clock frequencies to 20GHz \cite{rsfq}, instead of theoretically
predicted hundreds GHz. Therefore, investigation of possible ways
for suppression of time jitter of transmitting signals is a very
important problem from fundamental and technological points of
view. An RSFQ circuit in fact has long timing loops, so the time
jitter produces a noise-induced digital accumulative effect in
Josephson transmission lines \cite{rl,s}. It is also important for
different branches of physics, where a nonlinear element is driven
either from an external source or from another element, as in
neural networks. For RSFQ circuits three different types of
digital errors may be identified. First, storage errors may occur
during the passive storage of data: fluctuations can induce a
$2\pi$ phase flip and thus switch the quantizing loop into the
neighboring flux state. The analytical description of the mean
time of such noise-induced flips, valid for arbitrary noise
intensity, has been presented in \cite{MP}. Second, decision
errors may be produced in the two junction comparator. This
situation has been studied in detail in the linearized overdamped
JJ model \cite{compar}. If an RSFQ circuit operates at high speed,
another type of noise-induced error becomes important. Due to
noise the time interval between input and output pulses
fluctuates, producing "timing" error. Timing errors have also been
studied in the linearized overdamped JJ model \cite{rl}. It is
known that in noisy nonlinear systems with a metastable state, the
resonant activation (RA) \cite{Doe,PS,ACP} and noise enhanced
stability (NES) \cite{MP,MS} phenomena may be observed. These
effects may play positive and negative role in accumulation of
fluctuational errors in RSFQ logic devices: the RA phenomenon
minimizes timing errors, while the NES phenomenon increases the
switching time. These effects however were not still observed in
previous investigations due to the use of linearized models
\cite{rl,compar}. Moreover the limiting frequencies of RFSQ
devices and possible optimizations, in order to increase working
frequencies and reduce timing errors in RSFQ circuits, is an open
question.

This letter is aimed to answer this question by investigating
nonlinear noise properties of an overdamped JJ, subjected to
periodic driving, to understand possible ways of RSFQ circuits
optimization for high-frequency operation with minimal timing
errors. To this end we consider the dynamics of a short overdamped
JJ, under a current $I$, given by the Langevin
equation \cite{lik-B}
\begin{equation}
\omega _c^{-1}{\frac{d\varphi (t)}{dt}}=-{\frac{du(\varphi )}{%
d\varphi }} + i_F(t), \label{1}
\end{equation}
\begin{equation}
u(\varphi )=1-\cos \varphi -i(t)\varphi, \,\,\,\,\, i(t)=i_0+f(t),
\label{pot}
\end{equation}
where $\varphi $ is the order parameter phase difference,
$u(\varphi )$ is the dimensionless potential profile, $f(t)$ is
the driving signal, $i=I/I_c$ with $I_c$ the critical current, and
$i_F(t)=I_F/I_c$. $I_F$ is the thermal noise current with
$\left<I_F(t)\right>=0, and \left <I_F(t)I_F(t+\tau )\right
>={2\gamma }\delta (\tau )/{\omega _c}$. Here $\gamma =2ekT/(\hbar
I_c)$ is the dimensionless noise intensity, $T$ is the
temperature, $k$ is the Boltzmann constant,
$\omega_c=2eR_NI_c/\hbar$ is the characteristic frequency of the
JJ, $R_N$ is the normal state resistance of a JJ, $e$ is the
electron charge and $\hbar $ is the Planck constant. Let,
initially, the JJ is biased by a current smaller than the critical
one ($i_0<1$), and the junction is in the superconductive state.
The current pulse $f(t)$, such that $i(t)>1$, switches the
junction into the resistive state. However, an output pulse will
be born not immediately, but later on. Such a time is the
switching time, which is a random quantity characterized by mean
value and standard deviation. As an example of a driving with
sharp fronts we consider the dichotomous signal $f(t)=A {\rm
sign}(\sin(\omega t))$, and as an example of a driving with smooth
fronts -- sinusoidal signal $f(t)=A\sin(\omega t)$. In spite that
we consider the periodic driving, that makes the results more
evident, it is obvious, that below the cut-off frequency the
switching occurs during half of the first period, so other periods
are not important. Besides, our adiabatic analysis may easily be
generalized to an arbitrary form of $f(t)$. In computer
simulations we set $\omega_c=1$ and, therefore, in plots $\omega$
is normalized to $\omega_c$. The mean switching time (MST) $\tau =
\left< t \right > = \int_0^\infty t w(t)dt,$ and its standard
deviation (SD)  $\sigma$, may be introduced as characteristic time
scales of the evolution of the probability
$P(t)=\int_{\varphi_1}^{\varphi_2} W(\varphi,t)d \varphi$ to find
the phase within one period of the potential profile
Eq.(\ref{pot})
\begin{equation}
w(t) =  \frac{\partial P(\varphi_0,t)}{\partial
t[P(\varphi_0,\infty)-P(\varphi_0,0)]}. \label{prob}
\end{equation}
We choose therefore $\varphi_2=\pi$, $\varphi_1=-\pi$ and the
initial distribution at the bottom of a potential well.
\begin{figure}[th]
\epsfxsize=7.5cm \epsfbox{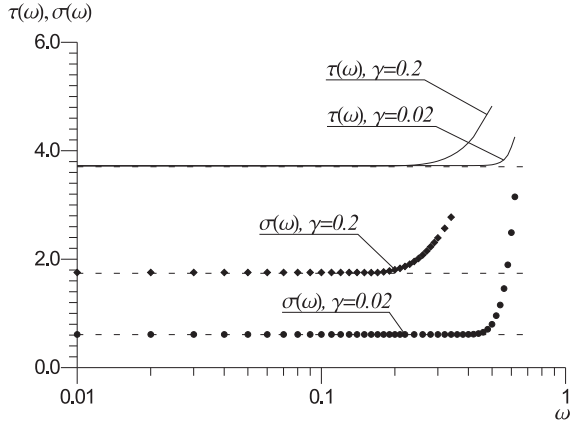} {\caption{ The MST
$\tau(\omega)$ and SD $\sigma(\omega)$ as functions of the
frequency for dichotomous driving, for two values of noise
intensity: $\gamma = 0.2, 0.02$, and $i_0=0.5$, $i=1.5$. The
results of computer simulations are: $\tau(\omega)$ (solid line)
and $\sigma(\omega)$ (diamonds and circles). Dashed line are the
theoretical results given by Eqs.(\ref{taucg}), (\ref{sg}).}
\label{fig1}}
\end{figure}
Let us first consider the case of dichotomous driving. From the
Fokker-Planck equation corresponding to the Langevin Eq.
(\ref{1}), we calculate the MST and its SD. In Fig. 1 we report
the behaviors of MST $\tau(\omega)$ and SD $\sigma(\omega)$ as a
function of the driving frequency for two values of noise
intensity: $\gamma = 0.2, 0.02$. For dichotomous driving both MST
and SD do not depend on the driving frequency below a certain
cut-off frequency, above which the characteristics degrade. In the
frequency range $0 \div 0.2 \omega_c$, therefore, we can describe
the effect of dichotomous driving by time characteristics in a
constant potential. The exact analytical expression of MST and its
asymptotic expansion are: for arbitrary $\gamma$ \cite{MP}
\begin{eqnarray}\nonumber
\tau_c(\varphi_0)&=&\displaystyle{\frac{1}{\gamma\omega _c}\left\{
\int_{\varphi_0}^{\varphi_2}e^{u(x)/\gamma }
\int_{\varphi_1}^xe^{-u(\varphi)/\gamma }
d\varphi dx+\right.}  \\
& & \displaystyle{\left.\int_{\varphi_1}^{\varphi_2}
e^{-u(\varphi)/\gamma}d\varphi\cdot \int_{\varphi_2}^\infty
e^{u(\varphi)/\gamma }d\varphi \right\}},\label{tauc}
\end{eqnarray}
and for $\gamma\ll 1$
\begin{eqnarray}
\tau_c(\varphi_0) = \frac{f_1(\varphi_2)-f_1(\varphi_0)+
 \gamma\left[f_2(\varphi_2)+f_2(\varphi_0) \right] + ..}{\omega _c},
 \label{taucg}
\end{eqnarray}
\begin{eqnarray}\label{f1}
f_1(x)&=&\frac 2{\sqrt{i^2-1}}
\arctan\left( \frac{i\tan(x/2)-1}{\sqrt{i^2-1}}\right), \\
\nonumber f_2(x)&=&1/({2(i-\sin{x})^2}).
\end{eqnarray}
Following ref. \cite{ACP}, the exact expression for
$\tau_{2c}=\left<t^2\right>$ in a time-constant potential may be
derived
\begin{eqnarray}
\tau_{2c}(\varphi_0)&=&\tau_c^2(\varphi_0)- \frac{2}{(\gamma\omega
_c)^2} \int_{\varphi_0}^{\varphi_2}e^{-u(x)/\gamma }
H(x)dx - \nonumber \\
& & \frac{2}{(\gamma\omega _c)^2}
H(\varphi_0)\int_{\varphi_1}^{\varphi_0} e^{-u(x)/\gamma}dx,
\label{tau2c}
\end{eqnarray}
where $H(x)=\int\limits_{x}^{\infty}e^{u(v)/\gamma }
\int\limits_{v}^{\varphi_2}e^{-u(y)/\gamma }
\int\limits_{y}^{\infty}e^{u(z)/\gamma } dzdydv$, and
$\tau_c(\varphi_0)$ is given by (\ref{tauc}). The asymptotic
expression of $\sigma=\sqrt{\tau_{2c}-\tau_c^2}$ in the small
noise limit $\gamma\ll 1$ is
\begin{eqnarray}\label{sg}
\sigma(\varphi_0)=\frac{1}{\omega_c}\sqrt{2\gamma
\left[F(\varphi_0)+f_3(\varphi_0)\right]+\dots },\\
\begin{array}{ccc}
F(\varphi_0)&=&f_1(\varphi_2)f_2(\varphi_2)-2f_1(\varphi_0)f_2(\varphi_0)+ \nonumber \\
& & f_1(\varphi_0)f_2(\varphi_0)
+\frac{f_1(\varphi_2)-f_1(\varphi_0)}{(i-\sin(\varphi_0))^2}, \nonumber \\
f_3(\varphi_0)&=& \int_{\varphi_0}^{\varphi_2}
\left[\frac{\cos(x)f_1(x)}{(\sin(x)-i)^3}-\frac{3}{2(\sin(x)-i)^3}
\right]dx.
\end{array}
\end{eqnarray}
The comparison between the asymptotic theoretical results
(Eqs.(\ref{taucg}),(\ref{sg})) and simulations is reported in Fig.
1. The agreement is very good within the frequency range $0 \div
0.2 \omega_c$. It is interesting to see that the SD of the
switching time scales as square root of noise intensity. The
dependencies of the MST and its SD on the bias current are
presented in Fig. 2 for $\gamma=0.001$. The agreement between
theoretical results and simulations is very good  for all the bias
current values investigated. In the low noise limit
Eq.(\ref{taucg}) gives actually the same results of ref.
\cite{rl}, since the largest contribution to the MST comes from
the deterministic term $\tau_c(\varphi_0)=\frac 1{\omega
_c}\left\{f_1(\varphi_2)-f_1(\varphi_0)\right\}$. However the Eq.
(\ref{sg}), in some cases, significantly deviates from the
linearized calculations. For $\gamma=0.001$ and $i=1.2$,
$\sigma=0.4 \omega_c^{-1}$ in ref.\cite{rl}, while we get
$\sigma=0.436 \omega_c^{-1}$. For larger current $i=1.5$, the
discrepancy is larger: $\sigma=0.06 \omega_c^{-1}$ in \cite{rl},
and we get $\sigma=0.14 \omega_c^{-1}$. In the inset of Fig. 2 we
report the behaviour of SD (Eq.(\ref{sg})) as a function of noise
intensity $\gamma$. For noise intensity values up to $\gamma =
0.05$ the agreement with computer simulations is very good.
Therefore not only low temperature ($\gamma \leq 0.001$), but also
high temperature devices may be described by Eqs.(\ref{taucg}),
(\ref{sg}). If noise intensity is rather large, the phenomenon of
NES may be observed: the MST increases with the noise intensity,
as it may be easily seen from Eq.(\ref{taucg}). In the design of
large arrays of RSFQ elements, operating at high frequencies, it
is very important to consider this effect, otherwise it may lead
to malfunctions due to the accumulation of digital errors.
\begin{figure}[th]
\epsfxsize=7.5cm \epsfbox{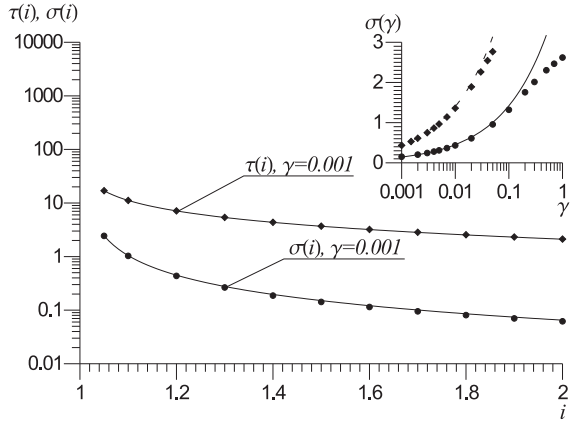} {\caption{ The MST and SD vs
bias current for the time-constant case for $\gamma=0.001$. Solid
lines - formulas (\ref{taucg}) and (\ref{sg}), diamonds and
circles - results of computer simulation. Inset: the SD vs noise
intensity for the time-constant case ($i>1$). Solid line - formula
(\ref{sg}), circles - results of computer simulation for $i=1.5$.
Dashed line - formula (\ref{sg}), diamonds - results of computer
simulation for $i=1.2$. } \label{fig2}}
\end{figure}

Now let us consider the case of sinusoidal driving. The
corresponding time characteristics may be derived using the
modified adiabatic approximation \cite{PS,ACP}
\begin{equation}
P(\varphi_0,t)=\exp\left\{-\int_{0}^t
\frac{1}{\tau_c(\varphi_0,t')}dt'\right\},
\label{met}
\end{equation}
with $\tau_c(\varphi_0,t')$ given by (\ref{tauc}). For dichotomous
driving the value of initial current $i_0$ has a weak effect on
temporal characteristics, as in ref. \cite{rl}, while it is
important for sinusoidal driving , since it also defines the
potential barrier height. We focus now on the current value
$i=1.5$, because $i=1.2$ is too small for high frequency
applications.
\begin{figure}[h]
\epsfxsize=7.5cm \epsfbox{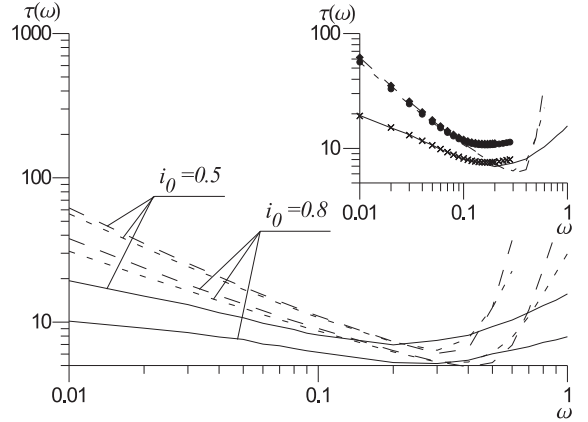} {\caption{ The MST vs
frequency for $f(t)=A\sin(\omega t)$ (computer simulations) for
$i=1.5$. Long-dashed line - $\gamma=0.02$, short-dashed line -
$\gamma=0.05$, solid line - $\gamma=0.5$, from top to bottom
$i_0=0.5;A=1$, $i_0=0.8;A=0.7$. Inset: comparison between
simulations and theoretical results obtained from Eq.(\ref{met})
for $i_0=0.5;A=1$, and $\gamma=0.02$ (diamonds), $\gamma=0.05$
(circles) and $\gamma=0.5$ (crosses). } \label{fig3}}
\end{figure}
In Fig. 3 the MST as a function of the driving frequency for
different values of bias current is shown. For smaller $i_0$ the
switching time is larger, since $\varphi_0=\arcsin(i_0)$ depends
on $i_0$. On the other hand, the bias current $i_0$ must be not
too large, since it will lead, in absence of driving, to the
reduction of the mean life time of superconductive state, i.e. to
increasing storage errors (Eq.(\ref{tauc})). Therefore, there must
be an optimal value of bias current $i_0$, giving minimal
switching time and acceptably small storage errors.
 Storage errors, in fact, are acceptably small up to $i_0=0.99$
 for $\gamma=0.001$ \cite{rl}.
 We observe the phenomenon of resonant activation: MST has a
minimum as a function of driving frequency. The approximation
(\ref{met}) does not describe the resonant activation effect at
high frequencies, but it works rather well below 0.1 $\omega_c$,
that is enough for practical applications. Moreover, it is
interesting to see that near the minimum the MST has a very weak
dependence on the noise intensity, i. e. in this signal frequency
range the noise is effectively suppressed. We observe also the NES
phenomenon. There is a frequency range, around $0.2-0.4 \omega_c$
for $i_0=0.5$ and around $0.3-0.5 \omega_c$ for $i_0=0.8$, where
the switching time increases with the noise intensity. The NES
effect increases for smaller $i_0$ because the potential barrier
disappears for a short time interval within the driving period $T
= 2\pi/\omega$ \cite{MS} and the potential is more flat, so noise
has more chances to prevent the phase to move down and delay
switching process. This effect may be avoided, if the operating
frequency does not exceed 0.2$\omega_c$. Besides (see Fig. 4) the
SD also increases above 0.2$\omega_c$.
\begin{figure}[th]
\epsfxsize=7.5cm \epsfbox{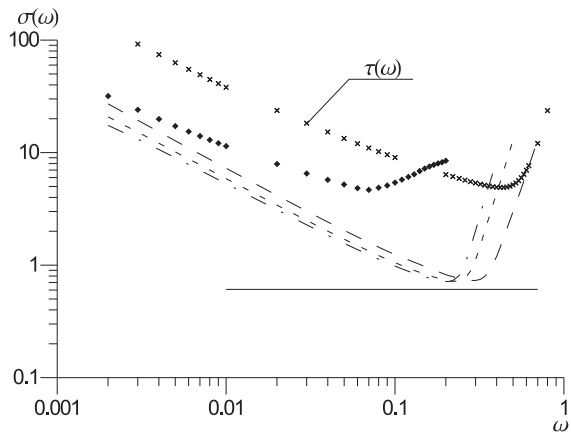} {\caption{ The SD vs frequency
for $f(t)=A\sin(\omega t)$ and $\gamma = 0.02$. Computer
simulations - dash-dotted line: $i_0=0.3;A=1.2$, short-dashed
line: $i_0=0.5;A=1$, long-dashed line: $i_0=0.8;A=0.7$. The MST is
given by crosses for comparison ($i_0=0.8;A=0.7$). Formula
(\ref{sg}) - solid line. Formula (\ref{met}) - diamonds. }
\label{fig4}}
\end{figure}
The plots of SD as a function of driving frequency for
$\gamma=0.02$, $i=1.5$ and different values of $i_0$
 are shown in Fig. 4. The approximation (\ref{met}) is
not so good for SD as for MST, even if the qualitative behaviour
of SD is recovered. We see that the minimum of $\sigma(\omega)$,
for $\gamma=0.02$, is located near the corresponding minimum for
$\tau(\omega)$ in Fig. 3. For the SD the optimal frequency range,
where the noise induced error will be minimal, is from $0.1$ to
$0.3$ for the considered range of parameters. It is interesting to
see that, near the minimum, the SD for sinusoidal driving actually
coincides with SD for dichotomous driving (Eq.(\ref{sg})). This
means that even in the case of smooth driving, the limiting value
of SD may nearly be reached, but the RSFQ circuit must be properly
optimized.

 Finally we note that close location of minima
of MST and its SD means that optimization of RSFQ circuit for fast
operation  will simultaneously lead to minimization of timing
errors in the circuit, which is the main result of this paper.

    In the present paper we reported an analytical and
numerical analysis of influence of fluctuations and periodic
driving on temporal characteristics of the JJ. For dichotomous
driving the analytical expression of standard deviation of
switching time works in practically interesting frequency range
and for arbitrary noise intensity. For sinusoidal driving the
resonant activation effect has been observed in the considered
system: mean switching time has a minimum as a function of driving
frequency. Near this minimum the standard deviation of switching
time takes also a minimum value. The RA phenomenon was observed
very recently in the underdamped Josephson tunnel junction
\cite{YH}. Our theoretical investigation could motivate
experimental work in overdamped JJs as well. Utilization of this
effect in fact allows to suppress time jitter in practical RSFQ
devices and, therefore, allows to significantly increase working
frequencies of RSFQ circuits. Our study is not only important to
understand the physics of fluctuations in a Josephson junction, to
improve the performance of complex digital systems, but also in
nonequilibrium statistical mechanics of dissipative systems, where
noise assisted switching between
metastable states takes place \cite{Doe,PS, ACP,MS,YH}.\\
 \indent The work
has been supported by INTAS (projects 01-0367 and 01-0450), MIUR,
INFM, by the RFBR (projects 02-02-16775, 03-02-16533 and
02-02-17517), by the grant SS-1729.2003.2, and by RSSF.


\begin{thebibliography}{99}


\bibitem{rsfq} K. K. Likharev and V. K. Semenov, IEEE Trans. Appl. Supercond.,
{\bf 1}, 3 (1991); M. Dorojevets, P. Bunyk and D. Zinoviev,
\textit{ibid}{\bf 11}, 326 (2001); P. Bunyk, K. Likharev, and D.
Zinoviev, Int. J. High Speed Electron. Syst. {\bf 11}, 257 (2001).

\bibitem{MSS-RF} Y. Makhlin, G. Schon and A. Shnirman, Rev. Mod. Phys.,
{\bf 73}, 357 (2001).

\bibitem{lik-B}  K. K. Likharev, {\it Dynamics of Josephson Junct. and
Circ.}, Gordon and Breach, 1986; A. Barone and G. Paterno, {\it
Physics and Appl. of the Josepson Effect}, Wiley, 1982.

\bibitem{rl} A. V. Rylyakov and K. K. Likharev, IEEE Trans. Appl. Supercond.,
{\bf 9}, 3539 (1999).

\bibitem{s} J. Satchell, IEEE Trans. Appl. Supercond.,
{\bf 9}, 3841 (1999); Q. Herr, M. Johnson and M. Feldman,
\textit{ibid} {\bf 9}, 3594 (1999); V. Kaplunenko and V.
Borzenets, \textit{ibid} {\bf 11}, 288 (2001); T. Ortlepp, H.
Toepfer and H. F. Uhlmann, \textit{ibid} {\bf 11}, 280 (2001).

\bibitem{MP} A. N. Malakhov, and A. L. Pankratov,
Physica C, {\bf 269}, 46 (1996).

\bibitem{compar} T. J. Walls, T. V. Filippov, and K. K. Likharev, Phys. Rev. Lett.,
{\bf 89}, 217004 (2002).

\bibitem{Doe} C.R. Doering, J.C. Gadoua, Phys. Rev. Lett.,
\textbf{69}, 2318 (1992); R.N. Mantegna, B. Spagnolo,
Phys. Rev. Lett., \textbf{84}, 3025 (2000).

\bibitem{PS}A. L. Pankratov, and M. Salerno, Phys.
Lett. A {\bf 273}, 162 (2000).

\bibitem{ACP} A. N. Malakhov, and A. L.
Pankratov, Adv. Chem. Phys. {\bf 121}, 357 (2002).

\bibitem{MS} R.N. Mantegna, and B. Spagnolo,
Phys. Rev. Lett. {\bf 76}, 563 (1996); N. V. Agudov, and A. N.
Malakhov, Phys. Rev. E {\bf 60}, 6333 (1999); N.V. Agudov, B.
Spagnolo, Phys. Rev. E \textbf{64}, 035102(R) (2001);
 N.V. Agudov, A.A. Dubkov, B. Spagnolo, Physica A
\textbf{325}, 144 (2003); A. Fiasconaro, D. Valenti, B. Spagnolo,
\textit{ibid} \textbf{325}, 136 (2003).

\bibitem{YH} Y. Yu and S. Han,
Phys. Rev. Lett. {\bf 91}, 127003 (2003).

\end{thebibliography}
\end{document}